# Evidence for Localized Moment Picture in Mn-based Heusler Compounds

J. Karel[1], F. Bernardi[2], C. Wang[1], A. Beleanu[1], R. Stinshoff[1], N.-O. Born[1], S. Ouardi[1], U. Burkhardt[1], G.H. Fecher[1], C. Felser[1]

[1]Max-Planck-Institut für Chemische Physik fester Stoffe, Dresden, Germany 01187
[2]Departamento de Fisica, Instituto de Fisica, Universidade federal do Rio Grande do Sul (UFRGS), Porto Alegre, RS, Brasil

**Abstract**

X-ray absorption spectroscopy (XAS) and X-ray magnetic circular dichroism (XMCD) were used to probe the oxidation state and element specific magnetic moments of Mn in Heusler compounds with different crystallographic structure. The results were compared with theoretical calculations, and it was found that in full Heusler alloys, Mn is metallic (oxidation state near 0) on both sublattices. The magnetic moment is large and localized when octahedrally coordinated by the main group element, consistent with previous theoretical work, and reduced when the main group coordination is tetrahedral. By contrast, in the half Heusler compounds the magnetic moment of the Mn atoms is large and the oxidation state is +1 or +2. The magnetic and electronic properties of Mn in full and half Heusler compounds are strongly dependent on the structure and sublattice, a fact that can be exploited to design new materials.

## I. Introduction

Heusler compounds are a remarkable class of materials, which exhibit the complete range of physical properties. For example: half-metallic ferromagnets[1,2], heavy Fermion systems[3], semiconductors[4,5], shape memory alloys[3], superconductors[6], and topological insulators[2,4] all can be found within this material class. The versatility is due in part to the large number of different elements that can be used to form the ternary compounds on a relatively simple lattice with high symmetry, which, in the chemical sense, contains multiple sublattices. These sublattices can be represented by a combination of NaCl or ZnS-type structures[3], thus emphasizing that both octahedral and tetrahedral coordination by the main group element are present in these compounds.[3,7]

The Mn-based Heusler compounds can be used to investigate the properties of individual sublattices because compounds can be synthesized with Mn on the different sublattices. The site occupations, symmetries and coordination by the main group element are highlighted in Table 1 for representative structures within the Heusler class of compounds. The $X_2YZ$ ternary full Heusler structure (prototype $Cu_2MnAl$) is cubic with the space group $Fm\bar{3}m$. The 4a and 4c Wyckoff positions have octahedral symmetry, while the 8c position has tetrahedral symmetry, as indicated. In these compounds, Kübler predicted that Mn when coordinated octahedrally by the main group element (4b position) exhibits a large localized magnetic moment of ~3.5-4 $\mu_B$ (Kübler Rule).[8] The Mn majority electrons form a delocalized band with the majority electrons of the X element; however the Mn minority spin states are pushed above the Fermi energy, meaning they are excluded from the Mn sites. This results in a localized magnetic moment formed completely by itinerant electrons.[8] A variation of this structure is the inverse Heusler structure (prototype $Li_2AgSb$), which has 4 inequivalent lattice sites, each with tetrahedral symmetry; the space group is $F\bar{4}3m$.

A quaternary compound (prototype LiMgPdSn) is formed with the same symmetry and space group when all four Wyckoff positions (4a-4d) of space group 216 are occupied by different atoms, as shown in Table 1 (Note that the main group element is kept in 4a). Moreover, the so-called half Heusler compounds are a variation on this structure with the 4c or 4d position vacant (prototype MgAgAs).

The coordination of Mn with respect to the main group element, leads to unique physical properties, many of which are promising for technology applications. For instance, $Mn_3Ga$ and $Mn_{3-x}Co_xGa$ have been proposed as potential materials for spin transfer torque devices.[9] Here, Mn is present on multiple sublattices, and the magnetic moments are antiparallel, guaranteeing a low total moment with a high Curie temperature.[9] Additionally, Mn when coordinated octahedrally by Si in $Co_2MnSi$ leads to full spin-polarization of the charge carriers[10] and consequently widespread applicability as an electrode material in tunneling magnetoresistance (TMR) devices.[11] Designing new materials for future device applications necessitates a general understanding of how Mn behaves on the individual chemical sublattices. We therefore probed the electronic and magnetic properties of Mn as a function of coordination with respect to the main group element in full and half Heusler compounds experimentally using X-ray absorption spectroscopy (XAS) and X-ray magnetic circular dichroism (XMCD) and theoretically with *ab initio* calculations. Both theory and experiment show that the Mn oxidation state indicates delocalized metallic-like behavior in the full Heusler compounds, while in the half Heusler materials the oxidation state is between +1 and +2. In both half and full Heusler alloys, the magnetic moment in the octahedral sites is large and localized (atomic-like), but the moment in the tetrahedral sites is reduced.

## II. Methods

$Co_2MnSi$ ($L2_1$), $Co_2MnAl$ ($L2_1$)[12], $Co_2Mn_{0.5}Fe_{0.5}Si$ (pseudo-ternary, $L2_1$), LiMnAs (Half Heusler, $C1_b$), NiMnSb (Half Heusler, $C1_b$), $Mn_2CoAl$ (inverse, $X_a$), $Mn_2CoGa$ (inverse, $X_a$), $Mn_2VAl$ ($L2_1$) and CoFeMnZ (Z=Ga, Ge, Si; quaternary) were investigated in the present work. All compounds except LiMnAs were prepared by arc melting stoichiometric quantities of the constituent elements in an Ar atmosphere. Ti is premelted as an oxygen getter to prevent oxidation of the ingot. LiMnAs was prepared by solid state reaction; details are given in [13]. X-ray diffraction and energy dispersive X-ray spectroscopy (EDX) were used to confirm the structure and composition, respectively. The Mn oxidation state was determined by XAS (X-ray absorption spectroscopy) of the Mn $K$-edge at room temperature in transmission mode. These measurements were performed at HASYLAB DORIS beamline A1 at DESY (Hamburg, Germany), beamlines 16A and 17C at the National Synchrotron Radiation Research Center (Hsinchu, Taiwan) and beamline XAFS1 (Brazilian Synchrotron Light Laboratory, Campinas, Brazil). Mn, MnO and $MnO_2$ were measured as standards for the different Mn valences. A linear background in the pre-edge and post-edge regions was subtracted from all $K$-edge absorption data, and then the pre-edge (post-edge) region was normalized to an edge-jump of one. If $E_0$ is the energy position of the Mn metal absorption edge (6538.7eV), then the energy shift for each sample was calculated from $E_A - E_0$, where $E_A$ is the energy value of the absorption edge. In general, the XAS spectrum at the Mn K-edge presents several features. In the case of a pre-edge peak, the second peak of the derivative curve is defined as the absorption edge[14].

X-ray absorption spectroscopy (XAS) and X-ray magnetic circular dichroism (XMCD) at the Mn $L_{3,2}$-edges were performed at beamline 11A at the National Synchrotron Radiation Research Center (Hsinchu, Taiwan) at room temperature ($Mn_2VAl$) and 80 K (NiMnSb) in magnetic fields of ±1 T. The background was removed by subtracting a two-step function to take into account photoelectrons promoted to the continuum; the remaining spectra was then normalized to the number of Mn 3$d$ holes. $I^P$ and $I^{AP}$ are defined as the absorption intensities when the photon angular momentum is parallel and anti-parallel to the magnetization of the sample, respectively. Then the XAS is defined as ($I^P + I^{AP}$)/2 and the XMCD as $I^P - I^{AP}$. The spin and orbital magnetic moments are calculated using the sum rules, based on the methods described in ref [15]; 4.5 was used for the number of Mn $3d$ holes, based on *ab initio* calculations. A correction factor of 1.5 was applied following reference [16] for the calculation of the Mn spin moments. XAS spectra and extracted Mn magnetic moments for CoFeMnZ (Z=Ga, Ge, Si) were reproduced with permission from [17].

*Ab initio* calculations of the magnetic moments and theoretical X-ray absorption spectra were performed. The electronic structure and spectra were calculated using the full-potential linearized augmented plane wave (FLAPW) code WIEN2k[18,19] with parameters as described in detail in Reference [20]. The exchange-correlation functional was evaluated within the generalised gradient approximation, using the Perdew–Burke–Ernzerhof parameterisation.[21] The muffin-tin radii ($R_{MT}$) were set automatically using the standard subroutine of WIEN2k. $R_{MT} \times k_{max} = 7$ was used for the number of plane waves and the expansion of the wave functions was set to $l = 10$ inside of the muffin tin spheres. The self-consistent calculations employed a 25×25×25 grid of *k* points in the full Brillouin zone for integration. Its reduction to the irreducible wedge of the Brillouin zone results in different numbers of *k* points depending on the symmetry of the crystal structure. The energy convergence criterion was set to $10^{-5}$ Ry and the charge convergence was set to 0.01 electrons.

To calculate the X-ray absorption spectra, a primitive 1×1×1 cubic super-cell with 16 atoms was used to account for the core-hole at the emitting atom. The energy is given by $E = E(N) - E(N^*)$, where $E(N)$ is the energy of the ground state with $N$ electrons and $E(N^*)$ is the energy of the excited state with one Mn 1*s* core electron promoted to the valence band. The size of the super-cell was confirmed to be large enough so as to not influence the shape of the spectra. It should be mentioned that more precise energies may be found when using Slater's transition rule with only 1/2 electron removed from the initial orbital. Atomic-type many-particle calculations were performed to explain some details of the Mn 2*p* states in the photon absorption. The multiplet calculations were performed using de Groot's program ctm4xas[22], which includes the effects of the crystal field and charge transfer. The details of the multiplet description and applied methods are given in References [22,23,24,25].

### III. Results
*XAS at the Mn K-edge*

Mn in CoFeMnZ (Z=Ga, Ge, Si), $Co_2MnSi$, $Co_2MnAl$ and $Co_2Mn_{0.5}Fe_{0.5}Si$ is octahedrally coordinated by the main group element. Additionally, NiMnSb and LiMnAs are half Heusler materials, where Mn is also octahedrally coordinated. LiMnAs exhibits the half Heusler structure at high temperatures and undergoes a phase transition at 768 K to a tetragonal structure.[13] On the other hand, $Mn_2CoAl$ and $Mn_2CoGa$ crystallize in the inverse Heusler structure, which has two different Mn sublattices, one with tetrahedral coordination and the other with octahedral coordination. All Mn in $Mn_2VAl$ are located in sites with tetrahedral coordination with respect to the main group element.

Figure 1 shows the experimental Mn *K*-edge XAS of the compounds investigated. Samples with some or all Mn in the tetrahedrally coordinated sites ($Mn_2$- compounds) show similar spectra (and therefore electronic structure) and exhibit a large intensity at the absorption edge. This intensity is also observed in the ternary (and pseudo ternary) $Co_2$- compounds. However, it is not as pronounced in the quaternary CoFeMnZ samples. In contrast, the half Heusler compounds show a shift in the absorption edge, indicating an increase in valence. These results can generally be attributed to the differences in local symmetry among the compounds, which heavily influence the spectra.[26,27]

The oxidation state was determined based on the energy shift from the $Mn^0$ absorption edge using the method in ref [14]. A linear relationship exists between this shift and the oxidation state (Kunzl's rule), which can qualitatively be understood as an increase in the binding energy of 1*s* electrons with increasing oxidation state due to the attractive potential of the nucleus and a reduction in the repulsive core Coulomb interaction. Figure 2 shows the calculated Mn oxidation states versus the Mn *K*-edge absorption shift for the samples investigated and the Mn, MnO and $MnO_2$ standards. Most of the compounds appear metallic with oxidation states near zero. Surprisingly, NiMnSb and LiMnAs have oxidation states of +1 and +1.6, respectively.

*XAS and XMCD at Mn $L_{3,2}$ edges*

A smaller subset of these samples with different structures and Mn in different lattice positions was examined with XAS at the Mn L-edge; these data are shown in Figure 3 for CoFeMnSi, $Co_2MnAl$, $Mn_2CoAl$, $Mn_2VAl$ and NiMnSb. The CoFeMnSi and $Co_2MnAl$ compounds exhibit features on the high energy side of the $L_3$ absorption edge (labeled A and B), which are indicative of multiplet splitting originating from electron localization.[28,29,30,31] This result is consistent with similar observations in $Rh_2MnGe$, $Co_2Mn_xTi_{1-x}Si$ and CoFeMnAl where Mn is also octahedrally coordinated with respect to the Z element.[17,28,32,33] In $Mn_2CoAl$, where Mn is in both the octahedrally and tetrahedrally coordinated positions, feature A is not present and feature B is significantly reduced, reflecting the influence of the itinerant nature of sites with tetrahedral coordination of the main group element. By contrast, the spectrum for NiMnSb displays evidence of multiplet splitting in the $L_2$ absorption edge, which further supports the observation of the oxidation state near +1.

The absorption spectrum for $Mn_2VAl$, where Mn is only tetrahedrally coordinated by the main group element, is consistent with previous reports[34] and metallic Mn, reflecting the oxidation state near zero derived from the *K*-edge absorption spectra. A shoulder is present in the pre-edge of the $L_3$ white line; to identify the origin of this structure two spectra, with either strong LSJ coupling or enhanced jj coupling, were calculated. For the LSJ coupling calculation of the 2p excitation, the Slater integrals were scaled to 90% of their value from the Hartree–Fock calculations. The spectra were broadened by 250 meV according to the experimental resolution. The lifetime broadening was varied over the spectrum, with larger values used for the $p_{1/2}$ or $L_3$ parts of the spectra. In jj coupling a crystal field with 10$Dq$=1 eV was used together with Slater integrals reduced to 1%. In both cases, LSJ and jj coupling, the spin orbit coupling parameter was kept at 100%. Figure 4a and 4b show the calculated Mn *L*-edge spectra for LSJ and jj coupling respectively. When considering enhanced LSJ coupling, additional peaks appear on the high energy side of the $L_3$ edge in addition to splitting in the $L_2$ white line. These features are present in other Mn *L*-edge absorption spectra examined here, for instance CoFeMnSi (Figure 4c), and arise from electron localization and a strong interaction between the core hole and final state.[17] We note that previous work has shown these features can also result from dipole transitions between a ground state that is mixed due to *p-d* charge transfer and the corresponding final states, however charge transfer was not included in the calculations performed here.[32] By contrast, the shoulder on the low energy side of the $L_3$ white line observed in $Mn_2VAl$ (Figure 4d) is reproduced in the calculations when the jj coupling term is increased, indicating it originates from stronger crystal field splitting. Additionally, the splitting between the $L_3$ and $L_2$ edges is reduced in the case of jj coupling (Figure 4b); experimentally a reduction in this splitting is also observed for $Mn_2VAl$ and NiMnSb (Figure 3) and was previously reported for NiMnSb[35], further indicating a stronger jj coupling in these materials.

The influence of different local environments on the magnetic moments was examined by calculating the moments from the XMCD spectra using the sum rules. Table 2 shows the experimental and theoretical total magnetic moments for the compounds investigated. $Mn_2VAl$ and $Mn_2CoAl$ (tetrahedrally coordinated with respect to Z) display a reduced magnetic moment in comparison to the other compounds where Mn is in sites with octahedral coordination with respect to the main group element. The magnetic moments for $Co_2MnSi$ and CoFeMnZ (Z=Ga, Ge, Si) are taken from references 36 and 17, respectively. The relatively low magnetic moments for CoFeMnGe, CoFeMnGa compared to the theoretically predicted values were attributed to chemical disorder.[17] Previous studies on $Co_2Mn_xTi_{1-x}Si$, $Co_2MnGe$, $Co_2MnGa_{1-x}Ge_x$ and $Rh_2MnGe$ have similarly reported large Mn magnetic moments when Mn is octahedrally coordinated.[17,32,33]

## V. Discussion

We first begin with the full Heusler compounds. When Mn is octahedrally coordinated by the main group element, it exhibits a low oxidation state (based on K-edge absorption), multiplet splitting (based on L-edge absorption) and a large magnetic moment. The fact that the absorption data shows at the same time metallic behavior and evidence of localization (multiplet splitting) is seemingly contradictory but is in fact entirely consistent with previous theoretical predictions.[8] Kübler found the Mn magnetic moment is large and highly localized but comprised of delocalized electrons.[8] The multiplet splitting observed in the *L*-edge absorption data indicates localization of the d states and is consistent with the observed large magnetic moment. By contrast, the metallic behavior of the K-edge absorption reflects the itinerant *p* electrons, which are hybridized with *p* states from the Z element. These results confirm the theoretical picture in which ferromagnetism between localized Mn moments is mediated by itinerant electrons, as in an RKKY interaction.[8,37,38] On the other hand, Mn with tetrahedral coordination with respect to the main group element has both a low oxidation state and magnetic moment.

In the half Heusler compounds, the situation differs. The Mn, present in sites octahedrally coordinated by the main group element, exhibits both a large magnetic moment and an oxidation state between +1-2. These results indicate that Mn is more atomic-like in these compounds and suggests similarities with the manganites, which although the magnetic behavior is very complex, can often be explained by atomic-like integer moments.[39] Additionally, we note a further similarity. Rare earth elements, with localized *f* electrons, populate the lattice sites with octahedral coordination in half Heusler compounds[3]; the only stable half Heusler compounds with transition metals in this octahedral site are those with Mn in that position.

Depending on the sublattice and structure, Mn displays a wide range of magnetic and electronic properties. Indeed this behavior should be considered intermediate between itinerant band ferromagnets and the atomic-like localized moments in manganites. The fact that within one compound there exists two Mn sublattices with different properties affords the opportunity to tailor materials for specific applications. For instance, it may be possible to realize a room temperature ferromagnetic semiconductor by utilizing Mn in the octahedral sites. Moreover, coupling between Mn on different sublattices, which exhibit different magnetic moments and behavior, may have interesting dynamical effects that can be exploited for all optical switching devices, similar to recent work in GdFeCo.[40]

## VI. Conclusion

Theoretical calculations, X-ray absorption and X-ray magnetic circular dichroism were used to determine the valance state and magnetic moments, respectively, for a series of Mn-containing Heusler compounds with Mn on different chemical sublattices. When Mn is tetrahedrally coordinated by the main group element, it exhibits properties similar to a metal, i.e. a valance state near 0 and a low magnetic moment. In the full Heusler alloys, the Mn magnetic moment when octahedrally coordinated by the main group element is large and localized but the oxidation state shows metallic behavior. In the half Heusler alloys, the oxidation state is between +1 and +2, and the magnetic moments are large suggesting more atomic-like behavior. These general rules can be used to tailor the magnetic and electronic properties for future applications in Mn-based Heusler compounds by carefully controlling the individual sublattices.

**Acknowledgements**

Measurements at the National Synchrotron Radiation Research Center (Hsinchu, Taiwan) were preformed under proposals 2013-2-105-1 and 2013-2-027-1, and measurements at DESY (Hamburg, Germany) were executed under project number II-20100384.

| $Fm\bar{3}m$, no. 225, L2$_1$ | | | |
|---|---|---|---|
| Wyckoff position | 4a | 4b | 8c |
| **Site Symmetry** | $O_h$ | $O_h$ | $T_d$ |
| Coordination with respect to main group element | - | Octahedral | Tetrahedral |
| Co$_2$MnAl (Full Heusler, X$_2$YZ) 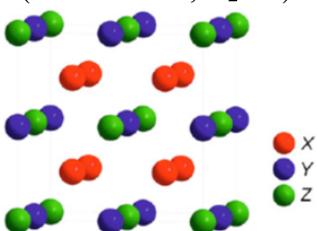 | Al | Mn | Co$_2$ |

| $F\bar{4}3m$, no. 216, C1$_b$ | | | | |
|---|---|---|---|---|
| Wyckoff position | 4a | 4b | 4c | 4d |
| **Site Symmetry** | $T_d$ | $T_d$ | $T_d$ | $T_d$ |
| Coordination with respect to main group element | - | Octahedral | Tetrahedral | Tetrahedral |
| Mn$_2$CoAl (inverse, X$_2$YZ) 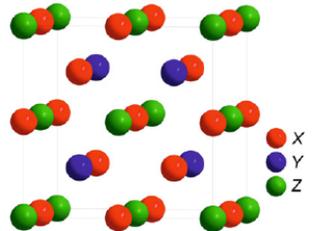 | Al | Mn | Mn | Co |
| CoFeMnAl (quaternary, XX'YZ) 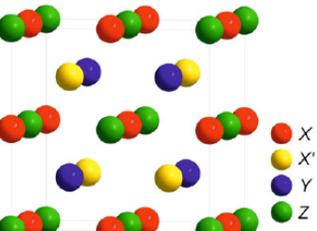 | Al | Mn | Fe | Co |
| MnNiSb (Half Heusler, XYZ) 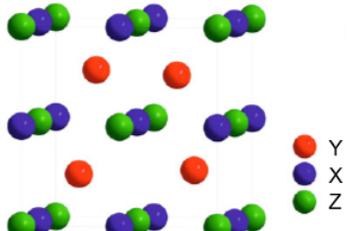 | Sb | Mn | Ni | Vacancy |

**Table 1.** Crystal Structures and site symmetries for Heusler compounds investigated in this work.

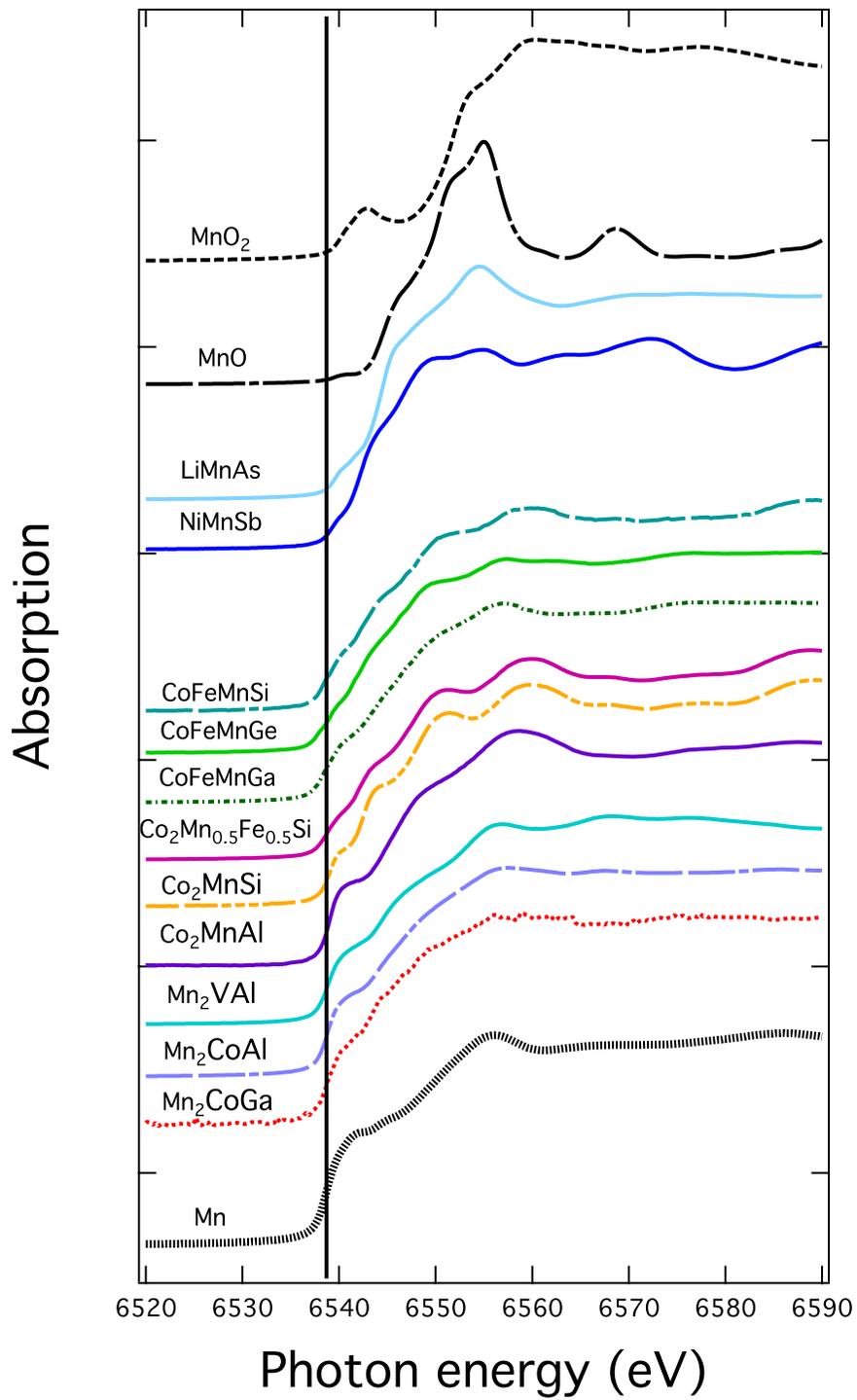

**Figure 1.** Mn K-edge X-ray absorption spectra.

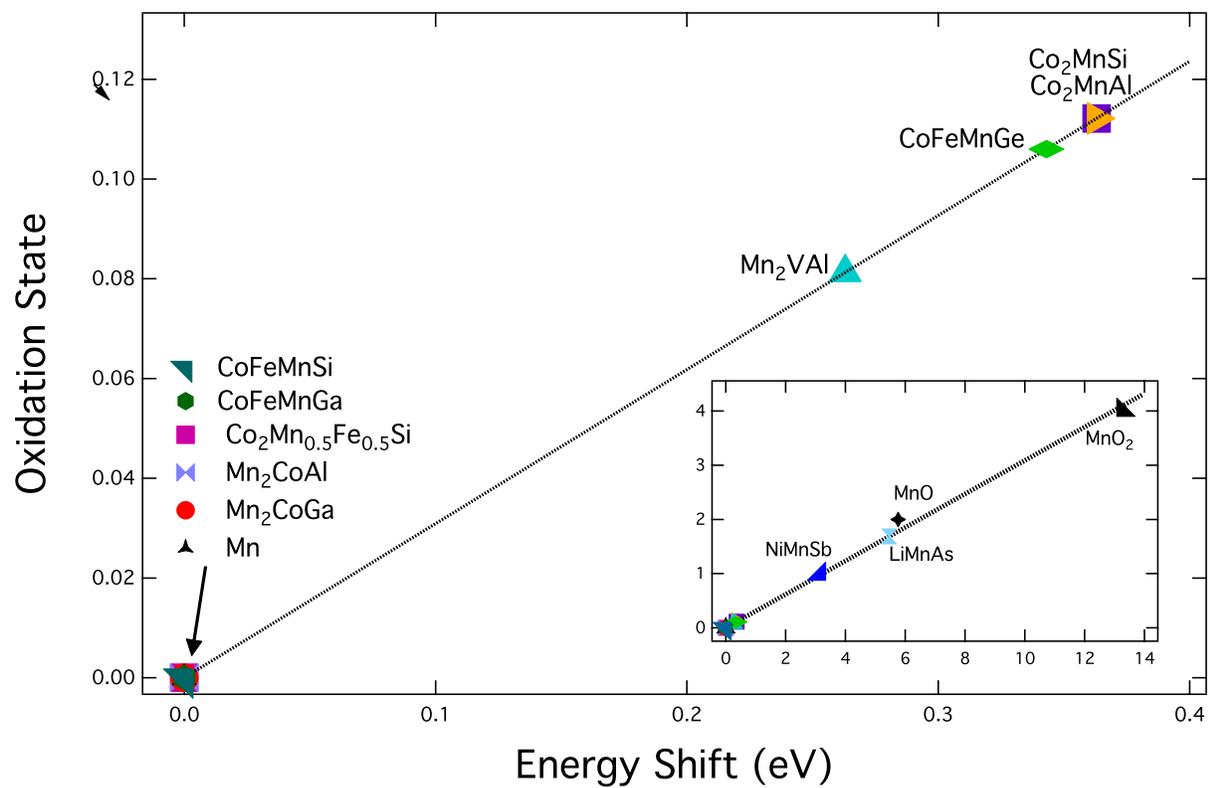

**Figure 2.** Oxidation state determined from Mn K-edge energy shift (as discussed in text). The inset shows an expanded range with $MnO_2$.

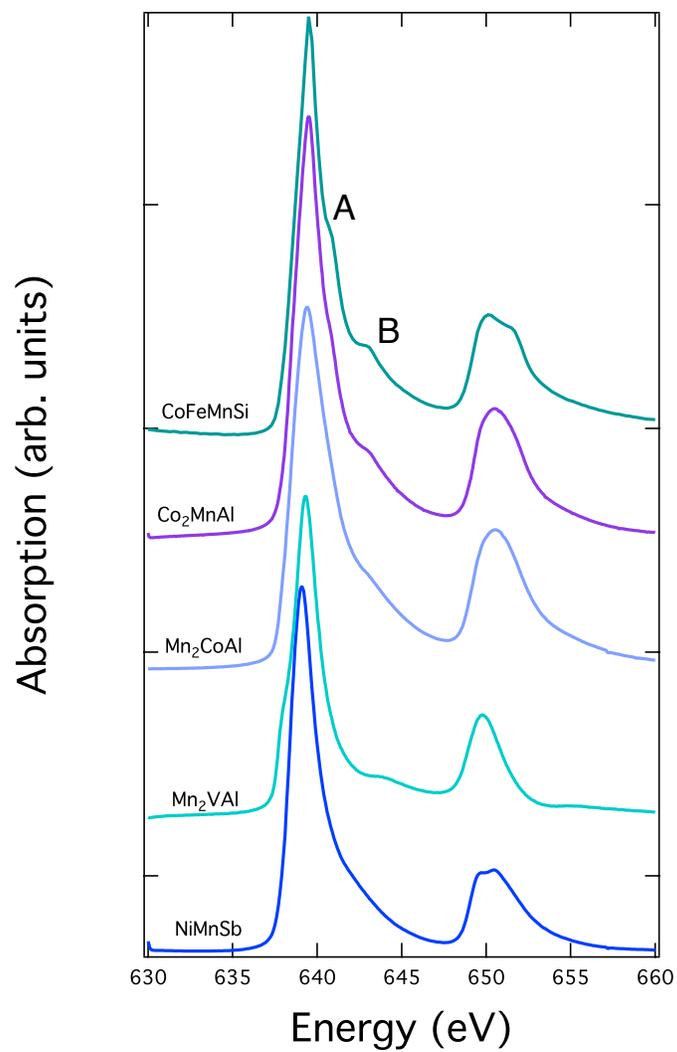

**Figure 3.** Mn *L*-edge absorption for selected compounds as indicated. CoFeMnSi spectrum is reproduced with permission from reference [17].

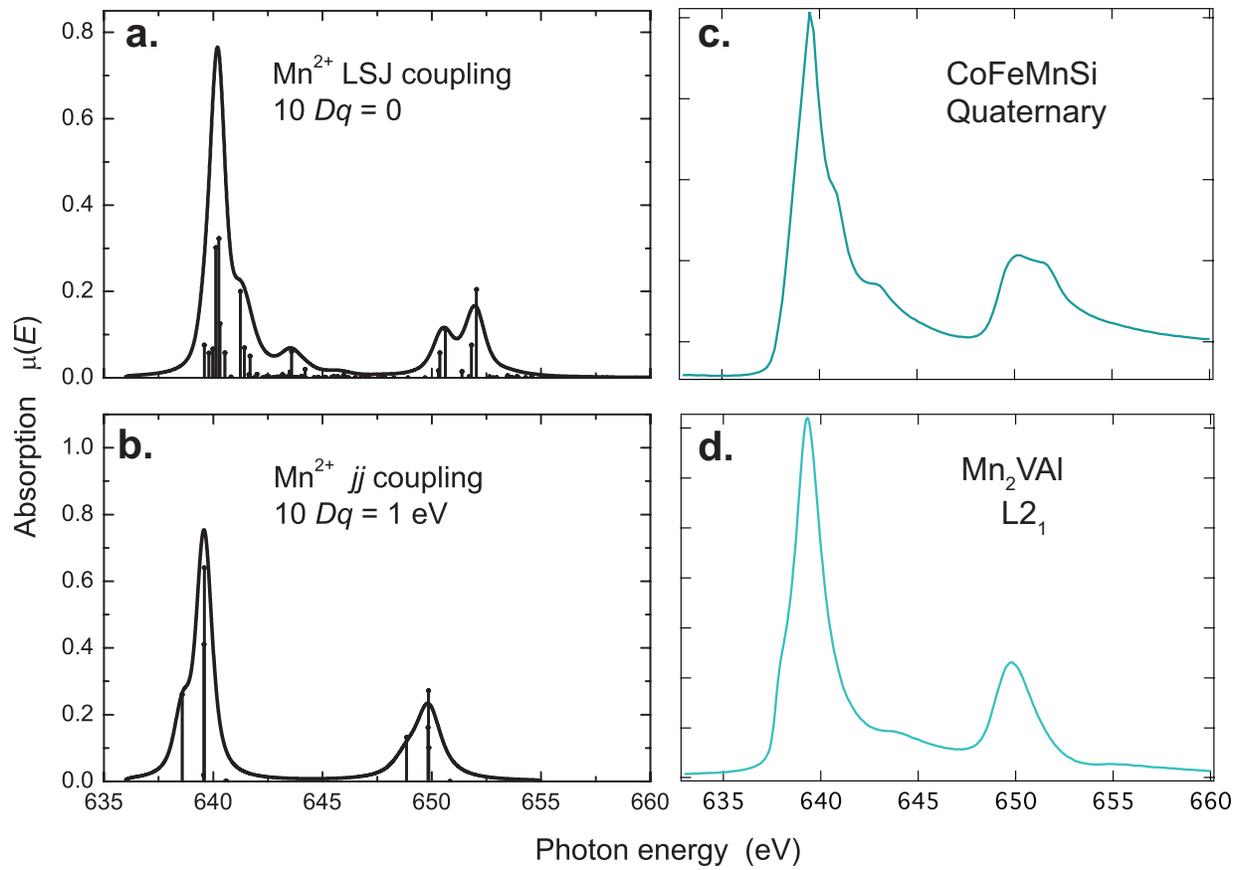

**Figure 4.** Calculated Mn *L*-edge absorption for **(a)** LSJ coupling and **(b)** jj coupling. In (a), the Slater integrals were scaled to 90% of their value from the Hartree–Fock calculations. In (b) a crystal field with $10Dq=1$ eV was used together with Slater integrals reduced to 1%. In both (a) and (b) the spin orbit coupling parameter was kept at 100%. Subfigures **(c) and (d)** show experimental Mn *L*-edge absorption of CoFeMnSi and $Mn_2VAl$, respectively for comparison

| Mn Coordination with respect to main group element | Compound | Experimental Mn Moment ($\mu_B$) | Calculated Mn Moment ($\mu_B$) |
|---|---|---|---|
| Octahedral | Co$_2$MnAl | | 2.67 |
| | Co$_2$MnSi* | 3.27 | 2.97 |
| | NiMnSb | 4.01 | 3.73 |
| | LiMnAs | | 4.15 |
| | CoFeMnGa** | 1.14 | 2.56 |
| | CoFeMnGe** | 1.99 | 2.72 |
| | CoFeMnSi** | 2.31 | 2.64 |
| Octahedral and Tetrahedral | Mn$_2$CoAl | | 2.62 (Mn$_I$) -1.53 (Mn$_{II}$) |
| Tetrahedral | Mn$_2$VAl | 1.97 | 1.40 |

**Table 2.** Calculated theoretical and experimental magnetic moments for tetrahedral and octahedral coordinated Mn atoms. Data marked with * is taken from ref [36] and data marked with ** from ref [17]. For comparison, reference [41] finds a Mn magnetic moment of 3.89 $\mu_B$ in LiMnAs.

# References

[1] R. Farshchi, M. Ramsteiner, *J. Appl. Phys.* **113** 191101 (2013)

[2] T. Graf, S.S.P. Parkin, C. Felser, *IEEE Trans. Magnet.* **47** 367 (2011)

[3] T. Graf, C. Felser and S.S.P. Parkin, *Prog. in Solid State Chem.* **39**, 1 (2011)

[4] F. Casper, T. Graf, S. Chadov, B. Balke, C. Felser, *Semicond. Sci. Technol.* **27** 063001 (2012)

[5] J.-W. G. Bos, R.A. Downie, *J. Phys. Condens. Matter* **26** 433201 (2014)

[6] T. Klimczuk, C. H. Wang, K. Gofryk, F. Ronning, J. Winterlik, G. H. Fecher, J.-C. Griveau, E. Colineau, C. Felser, J. D. Thompson, D. J. Safarik, and R. J. Cava, *Phys. Rev. B* **85** 174505 (2012)

[7] G. Kreiner, A. Kalache, S. Hausdor, V. Alijani, J.-F. Qian, G. Shan, U. Burkhard, S. Ouardi, C. Felser, *Z. Anorg. Allg. Chem.* **640**, 738 (2014)

[8] J. Kuebler, A.R. Williams, C.B. Sommers, *Phys. Rev. B* **28** 1745 (1983)

[9] J. Winterlink, S. Chadov, A. Gupta, V. Alijani, T. Gasi, K. Filsinger, B. Balke, G.H. Fecher, C.A. Jenkins, F. Casper, J. Kuebler, G-D. Liu, L. Gao, S.S.P. Parkin, C. Felser, *Adv. Mater.* **24** 6283 (2012)

[10] M. Jourdan, J. Minár, J. Braun, A. Kronenberg, S. Chadov, B. Balke, A. Gloskovskii, M. Kolbe, H.J. Elmers, G. Schönhense, H. Ebert, C. Felser, M. Kläui, *Nature Comm.* **5** 3974 (2014)

[11] X. Kozina, J. Karel, S. Ouardi, S. Chadov, G. H. Fecher, C. Felser, G. Stryganyuk, B. Balke, T. Ishikawa, T. Uemura, M. Yamamoto, E. Ikenaga, S. Ueda, K. Kobayashi, , *Phys. Rev. B* **89**, 125116 (2014)

[12] The presence of 4a/4b site disorder (B2 structure) could not be ruled out in this compound using the available Cu or Co $K_\alpha$ sources.

[13] A. Beleanu, J. Kiss, G. Kreiner, C. Kohler, L. Muchler, W. Schnelle, U. Burkhardt, S. Chadov, S. Medvediev, D. Ebke, and C. Felser G. Cordier and B. Albert A. Hoser, F. Bernardi, *Phys. Rev. B.,* **88** 184429 (2013)

[14] J. Wong, F.W. Lytle, R.P. Messmer, D.H. Maylotte, *Phys. Rev. B.,* **30**, 5596 (1984)

[15] C.T. Chen, Y.U. Idzerda, H.-J. Lin, N.V. Smith, G. Meigs, E. Chaban, G.H. Ho, E. Pellegrin and F. Sette, *Phys. Rev. Lett.* **75**, 152, (1995)

[16] Y. Teramura, A. Tanaka, T. Jo, *J. Phys. Soc. Jpn.* **65** 1053 (1996)

[17] P. Klaer, B. Balke, V. Alijani, J. Winterlink, G.H. Fecher, C. Felser, H.J. Elmers, *Phys. Rev. B*, **84**, 144413 (2011)

[18] K. Schwarz, P. Blaha, G.K.H.Madsen, *Comp. Phys. Comm.* **147** 71–76 (2002)

[19] P. Blaha, K. Schwarz, G. K. H. Madsen, D. Kvasnicka, and J. Luitz, WIEN2k: An Augmented PlaneWave + Local Orbitals Program for Calculating Crystal Properties (Wien, 2014).

[20] H.C. Kandpal, G.H. Fecher, C. Felser, *J. Phys. D* **40** 1507 (2007)

[21] J.P. Perdew, K. Burke, M. Ernzerhof, *Phys. Rev. Lett.* **77** 3865 (1996)

[22] E. Stavitski and F. M. F. de Groot. The CTM4XAS program for EELS and XAS spectral shape analysis of transition metal L edges. Micron, 41:687, 2010.

[23] R. D. Cowan. The Theory of Atomic Structure and Spectra. California Press, Berkeley, 1981.

[24] F. M. F. de Groot. Multiplett effects in X-ray spectroscopy. Coordin. Chem. Rev., 249:31, 2005.

[25] F. de Groot and A. Kotani. Core Level Spectroscopy of Solids. CRC Press, 2008.

[26] M. Croft, D. Sills, M. Greenblatt, C. Lee, S.-W. Cheong, K.V. Ramanujacharty, D. Tran, *Phys. Rev. B.* **55** 8726 (1997)

[27] F. Bridges, C.H. Booth, M. Anderson, G.H. Kwei, J.J. Neumeier, J. Snyder, J. Mitchell, J.S. Gardner, E. Brosha, *Phys. Rev. B.* **63** 214405 (2001)

[28] P. Klaer, B. Balke, V. Alijani, J. Winterlink, G.H. Fecher, C. Felser, H.J. Elmers, *Phys. Rev. B*, **84**, 144413 (2011)

[29] P. Klaer, M. Kallmayer, H.J. Elmers, L. Basit, J. Thoene, S. Chadov, C. Felser, *J. Phys. D.: Appl. Phys,* **42,** 084001 (2009)

[30] F. de Groot, *Coord. Chem. Rev.*, **249** 31 (2005)

[31] N.D. Telling, P.S. Keatley, G. van der Laan, R. J. Hicken, E. Arenholz, Y. Sakuraba, M. Oogane, Y. Ando, K. Takanashi, A. Sakuma, T. Miyazaki, *Phys. Rev. B*. **78** 184438 (2008)

[32] P. Klaer, M. Kallmeyer, C.G.F Blum, J. Barth, B. Balke, G.H. Fecher, C. Felser and H.J. Elmers, *Phys. Rev. B*, **80** 1444405 (2009)

[33] P. Klaer, T. Bos, M. Kallmayer, C.G.F. Blum, J. Barth, B. Balke, G.H. Fecher, C. Felser, H.J. Elmers, *Phys. Rev. B*, **82** 104410 (2010)

[34] P. Klaer, C.A. Jenkins, V. Alijani, J. Winterlik, B. Balke, C. Felser, H.J. Elmers, *Appl. Phys. Lett.* **98** 212510 (2011)

[35] I. Galanakis, S. Ostanin, M. Alouani, H. Dreysse, J.M. Wills, *Phys. Rev. B.,* **61** 4093 (2000)

[36] B. Krumme, D. Ebke, C. Weis, S. I. Makarov, A. Warland, A. Hueutten, and H. Wende, *Appl. Phys. Lett.* **101**, 232403 (2012)

[37] G. Malmstrom, D.J.W. Geldart, C. Blomberg, *J. Phys. F.:Metal Phys.* **6,** 233 (1976)

[38] G. Malmstrom, D.J.W. Geldart, C. Blomberg, *J. Phys. F.:Metal Phys.* **10,** 1953 (1976)

[39] J.M.D. Coey, M. Viret, S. von Molnar, *Adv. in Phys.* **48**, 167 (1999)

[40] C.E. Graves, A.H. Reid, T. Wang, B. Wu, S. de Jong, K. Vahaplar, I. Radu, D.P. Bernstein, M. Messerschmidt, L. Mueller, R. Coffee, M. Bionta, S.W. Epp, R. Hartmann, N. Kimmel, G. Hauser, A. Hartmann, P. Holl, H. Gorke, J.H. Mentink, A. Tsukamoto, A. Fognini, J.J. Turner, W.F. Schlotter, D. Rolles, H. Soltau, L. Strueder, Y. Acremann, A.V. Kimel, A. Kirilyuk, Th. Rasing, J. Stoehr, A.O. Scherz and H.A. Dürr, *Nature Mater.* **12** 293 (2013)

[41] P. J. Brown, A. P. Gandy, R. Kainuma, T. Kanomata, T. Miyamoto, M. Nagasako, K. U. Neumann, A. Sheikh, K.R.A. Ziebeck, *J. Phys. Cond. Matt.* **22** 206004 (2010)